\documentclass[reprint,nofootinbib,amsmath,amssymb,aps,prd,superscriptaddress]{revtex4-2}

\pdfoutput=1

\usepackage[T1]{fontenc}

\usepackage{graphicx}
\usepackage[dvipsnames]{xcolor}

\usepackage[hidelinks]{hyperref}
\hypersetup{colorlinks=true,linkcolor=Blue,citecolor=Blue,urlcolor=Blue}

\usepackage{amsmath}
\usepackage{amssymb}

\usepackage{amsthm}
\newtheorem{result}{Result}
\newtheorem*{corollary*}{Corollary}

\usepackage{calrsfs}

\newcommand{\myskip}{\hspace{9pt}}
\newcommand{\mybigskip}{\hspace{18pt}}

\DeclareMathAlphabet{\pazocal}{OMS}{zplm}{m}{n}
\usepackage[scr=boondox]{mathalfa}

\newcommand{\bigo}{\mathcal{O}}
\newcommand{\dif}{\mathrm{d}}

\newcommand{\T}{\mathrm{T}}
\newcommand{\eff}{\mathrm{eff}}
\newcommand{\e}{\mathrm{e}}
\newcommand{\iu}{\mathrm{i}}
\newcommand{\fo}{f(R_0)}
\newcommand{\fpo}{f'(R_0)}
\newcommand{\fppo}{f''(R_0)}
\newcommand{\barh}{\bar{h}}
\newcommand{\Rh}{{(h)}}
\newcommand{\szero}{{(0)}}
\newcommand{\sone}{{(1)}}
\newcommand{\stwo}{{(2)}}
\newcommand{\sboth}{{(1,2)}}
\newcommand{\srevboth}{{(2,1)}}
\newcommand{\Hone}{H^\sone}
\newcommand{\Htwo}{H^\stwo}
\newcommand{\Hboth}{H^\sboth}
\newcommand{\Hrev}{H^\srevboth}
\newcommand{\Do}{\nabla}
\newcommand{\Left}{\mathopen{}\mathclose\bgroup\left}
\newcommand{\Right}{\aftergroup\egroup\right}

\begin{document}

\title{Propagating degrees of freedom on maximally-symmetric backgrounds in $f(R)$ theories of gravity}

\author{Adri\'{a}n Casado-Turri\'{o}n}
    \email{adricasa@ucm.es}
    \affiliation{Departamento de F\'{i}sica Te\'{o}rica and Instituto IPARCOS, Universidad Complutense, 28040 Madrid, Spain}
    \affiliation{Fundamentini\k{u} Tyrim\k{u} Skyrius, Fizini\k{u} ir Technologijos Moksl\k{u} Centras (FTMC), 10257 Vilnius, Lithuania}
\author{\'{A}lvaro de la Cruz-Dombriz}
    \email{alvaro.dombriz@usal.es}
    \affiliation{ Departamento de F\'{i}sica Fundamental, Universidad de Salamanca, 37008 Salamanca, Spain}
    \affiliation{Cosmology and Gravity Group, Department of Mathematics and Applied Mathematics, University of Cape Town, Rondebosch 7700, Cape Town, South Africa}
\author{Antonio Dobado}
    \email{dobado@fis.ucm.es}
    \affiliation{Departamento de F\'{i}sica Te\'{o}rica and Instituto IPARCOS, Universidad Complutense, 28040 Madrid, Spain}

\date{\today}

\begin{abstract}
    In the context of $f(R)$ gravity, as well as other extended theories of gravity, the correct counting of globally well-defined dynamical modes has recently drawn a vivid interest. In this communication we present a consistent approach shedding light on such issues for both so-called degenerate and non-degenerate $f(R)$ models embedded in maximally-symmetric backgrounds. We find that the linearised spectrum of degenerate models on these backgrounds is empty, lacking both the graviton and scalaron modes which appear in generic non-degenerate models. Our work generalises previous results in the literature applicable only to the specific (degenerate) model $f(R)=\alpha R^2$; in fact, we find that the same pathologies discovered therein emerge for all choices of $f(R)$ belonging to the wide class of degenerate models.
\end{abstract}

\maketitle

\section{Introduction}

As widely known, out of all extensions of the theory of General Relativity (GR), $f(R)$ gravity \cite{Buchdahl:1970ynr,Sotiriou:2008rp,DeFelice:2010aj} stands out as the simplest generalisation thereof one could possibly envision. The ability of some particular $f(R)$ gravity models to successfully describe inflation \cite{Starobinsky:1980te} as well as the entire cosmic evolution \cite{Hu:2007nk,Nojiri:2006gh,Nojiri:2006be,Evans:2007ch}, together with the purported compliance with gravitational-wave observations \cite{Ezquiaga:2017ekz} have turned $f(R)$ theories into some of the most successful and observationally-viable alternative descriptions of gravity beyond the Einsteinian paradigm.

In spite of the aforementioned simplicity, the exact number of gauge-invariant, propagating degrees of freedom on physically-relevant backgrounds in $f(R)$ gravity has remained a matter of debate for several years, even when concentrating on the flat Minkowski background pertinent for studies on gravitational waves. Given that $f(R)$ theories are dynamically equivalent---under certain conditions, in the so-called Einstein frame---to GR plus a scalar field, dubbed the \emph{scalaron}, it was widely thought that, \emph{in vacuo}, the linear spectrum of $f(R)$ theories on a flat background would consist of the familiar massless and traceless graviton already found in GR, alongside with the additional massive scalar mode. In fact, this was the result found in pioneering investigations on the issue, such as \cite{Capozziello:2008rq}. This picture was later on put into question by studies claiming that the spectrum featured a second scalar mode, the \emph{breathing mode} \cite{Alves:2009eg,Alves:2010ms,RizwanaKausar:2016zgi}. Nonetheless, subsequent rigorous analyses \cite{Liang:2017ahj,Moretti:2019yhs} ended up refuting the existence of such a breathing mode.

Amidst these debates, \'{A}lvarez-Gaum\'{e} \textit{et al.}~were able to show \cite{Alvarez-Gaume:2015rwa} that the graviton was not present on the linearised spectrum of the particular model $f(R)=\alpha R^2$ atop its natural Minkowski background. In particular, they found that, upon linearisation of said model, the graviton perturbation lacked its corresponding kinetic term. This implies that the purely-quadratic $f(R)$ model features a strongly-coupled flat background, which is thus unstable. The existence of such an instability in an $f(R)$ model can be regarded as surprising, given that one of the key reasons behind the popularity of $f(R)$ theories was, precisely, their capability of avoiding the so-called Ostrogradski instabilities \cite{Ostrogradsky:1850fid} in spite of their fourth-order equations of motion. In fact, the presence of instabilities on physically-relevant backgrounds---such as Minkowski space-time or the cosmological Friedmann-Lema\^{i}tre-Robertson-Walker metric---has proven to be an instrumental criterion in assessing the viability of a given modified gravity theory. For some recent works pursuing this line of research, consult, for instance, \cite{BeltranJimenez:2019hrm,BeltranJimenez:2020lee,Jimenez-Cano:2022sds,Gomes:2023tur,Bello-Morales:2024vqk}; for an introductory review on the various kinds of instabilities appearing in field theories (including gravity theories), we refer the reader to \cite{Delhom:2022vae}.\footnote{
    In this communication, we shall employ the definitions of \emph{strong-coupling} and \emph{tachyonic} instabilities provided on reference \cite{Delhom:2022vae}.
}

Extending the aforementioned results by \'{A}lvarez-Gaum\'{e} \textit{et al.}, a proof that Minkowski space-time is strongly coupled in \emph{all} $f(R)$ models such that $f'(0)=0$ and $f(0)=0$ was provided years later in \cite{Casado-Turrion:2023rni}; this general proof covers the particular case $f(R)=\alpha R^2$. What is more, it was explicitly verified therein that $f(R)$ models fulfilling $\fpo=0$ for some constant scalar curvature $R_0$ are inherently pathological for several reasons. For instance, any background with $R=R_0\neq 0$ turns out to be unstable under perturbations of its Ricci scalar in models such that $\fpo=0$, while the stability of all backgrounds having $R_0=0$ in models with $f'(0)=0$ requires imposing a number of additional conditions on function $f$, some of which might be in contradiction with well-established viability criteria for $f(R)$ gravities, such as the Dolgov-Kawasaki stability condition \cite{Dolgov:2003px}. Because the field equations of all $f(R)$ models having $\fpo=0$ are trivially solved by every metric having constant scalar curvature $R=R_0$ (including those constant-curvature space-times featuring unphysical and pathological traits), said models were given the name \emph{$R_0$-degenerate} in \cite{Casado-Turrion:2023rni}.

Shortly afterwards, Hell, L\"{u}st and Zoupanos concluded \cite{Hell:2023mph} that the scalaron did not propagate either on top of Minkowski space-time in $f(R)=\alpha R^2$ gravity, adding to the non-propagation of the graviton already discovered in \cite{Alvarez-Gaume:2015rwa}. These results were later refined by Golovnev \cite{Golovnev:2023zen}, who explained such a phenomenology within the more general context of singular phase-space hypersurfaces in generic field theories. All in all, the aforementioned works evince that some aspects in the propagation and stability of degrees of freedom in $f(R)$ gravity are not fully understood, even on physically well-motivated backgrounds, such as Minkowski.

In the present work, we shall generalise the foregoing results, proving that maximally-symmetric backgrounds with $R=R_0$ are strongly coupled in \emph{every} $R_0$-degenerate $f(R)$ model, as well as in any model---degenerate or not---such that $\fppo=0$, with $R_0$ being a given constant-curvature solution of the theory. In fact, we shall show that the former models lack both the massless and traceless graviton as well as the massless scalaron when perturbed around a maximally symmetric background, whereas the latter models do not contain the scalaron only. We will also see that, by contrast, the three independent degrees of freedom corresponding to the graviton and the scalaron do propagate (stably) on maximally-symmetric backgrounds with $R=R_0$ in $f(R)$ models satisfying $\fpo\neq 0$ and $\fppo\neq 0$. Our findings---concisely summarised in Results \ref{result:DOF number ND}, \ref{result:graviton stability ND}, \ref{result:scalaron stability ND}, \ref{result:R_0 = 0 D stability} and \ref{result:R_0 neq 0 D stability} below---intend to complete the earlier partial findings already reported in the literature, settling the issue of the counting of propagating degrees of freedom on top of maximally-symmetric backgrounds in the context of $f(R)$ metric gravity once and for all.

The article shall be organised as follows. In Section \ref{Sec2} we provide a quick overview on maximally-symmetric space-times within the context of metric $f(R)$ gravities, while also reminding the reader of the precise notion of \text{$R_0$-(non-)degenerate} $f(R)$ models, which we shall then employ throughout this investigation. Afterwards, in Section \ref{Sec3} we obtain the linearised $f(R)$ equations of motion around such maximally-symmetric space-times, in such a way that they are valid for all $f(R)$ models, including those being degenerate. Once these equations are at hand, we turn our attention in Section \ref{sec:non-degenerate} to the study of the phenomenology of degrees of freedom for non-degenerate $f(R)$ models. Therein, we have devoted Subsections \ref{sec:graviton ND} and \ref{sec:scalaron ND} to the study of the graviton and the scalaron perturbations, respectively. Next, in Section \ref{Sec5} we analyse such degrees of freedom but for degenerate $f(R)$ models instead, with Subsections \ref{Sec5a} and \ref{Sec5b} respectively tackling the cases where the maximally-symmetric background possesses either zero or non-zero scalar curvature. Finally, we collect all our conclusions and prospects in Section \ref{Sec6}. Also, for the interested reader, Appendix \ref{Appendix: gauge-fixing} is devoted to shedding some light on gauge-fixing technicalities, Appendix \ref{Appendix: mode solutions} contains the derivation of mode solutions to the perturbation equations in a maximally-symmetric background of positive scalar curvature (de Sitter space-time), Appendix \ref{Appendix: m2 < 0} is dedicated to clarifying the physical meaning of models where the square of the scalaron mass is negative, and Appendix \ref{appendix:scalaron} features an explanation of the relationship between the Einstein-frame scalaron and the scalaron mode found in perturbation theory.

We recommend the busy reader to skip the detailed mathematical derivations in Subsections \ref{sec:graviton ND} and \ref{sec:scalaron ND} and Appendices \ref{Appendix: gauge-fixing} and \ref{Appendix: mode solutions}, and to concentrate directly on the Results \ref{result:graviton stability ND} and \ref{result:scalaron stability ND} derived from them.

\section{Maximally-symmetric backgrounds in $f(R)$ gravity}
\label{Sec2}

$f(R)$ theories are defined starting from the action
\begin{equation} \label{eq:f(R) action}
    S=\dfrac{1}{2 \kappa} \int \dif^4 x\,\sqrt{-g}\,f(R),
\end{equation}
where $\kappa \equiv 8 \pi G$. Hereafter, we shall employ the signature convention denoted as $(+,+,+)$ by Misner, Thorne and Wheeler \cite{Misner:1973prb}.\footnote{
    Notice that this signature convention differs from that employed in our previous work, \cite{Casado-Turrion:2023rni}.
} Variation of \eqref{eq:f(R) action} with respect to the metric provides the following set of equations of motion (EOM) in vacuum:
\begin{equation} \label{EOM}
    f'(R) R_{\mu\nu}-\dfrac{f(R)}{2}g_{\mu\nu}-(\nabla_\mu \nabla_\nu - g_{\mu\nu}\Box) f'(R)=0,
\end{equation} 
where primes denote differentiation with respect to $R$.

In this work we are interested in the phenomenology of $f(R)$ theories when small metric fluctuations around a maximally-symmetric (MS) space-time are performed. As widely known, such MS backgrounds are either de Sitter (dS), anti-de Sitter (AdS) or Minkowski space-times. This entails that MS backgrounds are endowed with a metric $\smash{g^\szero_{\mu\nu}}$ possessing a constant scalar curvature $R_0$, as well as with a Riemann tensor whose components satisfy
\begin{equation} \label{eq:Riemann MS}
    R_{\rho\mu\sigma\nu}^\szero=\dfrac{R_0}{12}\left[g_{\rho\sigma}^\szero g_{\mu\nu}^\szero-g_{\rho\nu}^\szero g_{\mu\sigma}^\szero\right].
\end{equation}
As such, one also has
\begin{equation} \label{eq:MS Ricci}
    R_{\mu\nu}^\szero=R^{\szero\rho}{}_{\mu\rho\nu}=\dfrac{R_0}{4}g_{\mu\nu}^\szero\equiv\Lambda g_{\mu\nu}^\szero,
\end{equation}
where $\Lambda$ plays the role of a cosmological constant.

When evaluated on a space-time with constant Ricci scalar $R_0$, such as the MS space-times, the vacuum $f(R)$ EOM \eqref{EOM} become
\begin{equation} \label{eq:EOM MS}
    \fpo R^\szero_{\mu\nu}=\dfrac{\fo}{2}g^\szero_{\mu\nu},
\end{equation}
whose trace reads
\begin{equation} \label{eq:EOM trace MS}
    \fpo R_0=2\fo.
\end{equation}
There are two distinct ways in which MS space-times can solve equations \eqref{eq:EOM MS} and \eqref{eq:EOM trace MS}, depending on the choice of function $f$:
\begin{itemize}
    \item If $\fpo\neq 0$, equations \eqref{eq:EOM MS} and \eqref{eq:EOM trace MS} are solved by MS space-times provided that their associated (effective) cosmological constant $\Lambda$ satisfies
    \begin{equation} \label{eqn:Lambda ND}
        \Lambda = \dfrac{R_0}{4} = \dfrac{\fo}{2\fpo}.
    \end{equation}
    In the following, we shall refer to these $f(R)$ models having $\fpo\neq 0$ as \emph{$R_0$-non-degenerate}, since their constant-curvature solutions---including those being MS space-times---are exactly the same as in GR $+$ $\Lambda$; this is a well-known result, cf.~\cite{Barrow:1983rx,Capozziello:2007id,delaCruz-Dombriz:2009pzc}, for instance. Notice that, in this scenario, $f(0)=0$ is a necessary condition when considering constant-curvature solutions with $R_0=0$, according to \eqref{eqn:Lambda ND}.
    \item If, on the contrary, there is a certain value of $R_0$ such that $\fpo=0$ (entailing, as per \eqref{eq:EOM trace MS}, that $\fo=0$), then \emph{any} constant-curvature metric with $R=R_0$---and, in particular, the corresponding MS space-time having scalar curvature $R_0$---becomes a trivial solution of the $f(R)$ model in question. These special models are therefore to be named \emph{$R_0$-degenerate}, as previously done in \cite{Casado-Turrion:2023rni}.
\end{itemize}
It should be stressed that some $f(R)$ models admit constant-curvature solutions with distinct Ricci scalars, hence the need for specifying a particular value of $R_0$ when discussing the degeneracy of a given model. In fact, some $f(R)$ models may be degenerate only for some---but not all---of their allowed constant-curvature solutions. This is the case, for instance, of the purely-quadratic model $f(R)=\alpha R^2$, which admits constant-curvature solutions for every $R_0$, but is only non-degenerate for $R_0\neq 0$, i.e.~it is $(R_0=0)$-degenerate. Another simple (yet illustrative) example featuring both degenerate and non-degenerate behaviour for two distinct values of $R_0$ is discussed extensively on Appendix C in \cite{Casado-Turrion:2023rni}. In those cases where it is absolutely clear from context that we are referring to a specific constant-curvature solution having $R=R_0$, we might refer to its host $R_0$-(non-)degenerate model as being simply \emph{(non-)degenerate}, so as to alleviate the terminology.

In what follows, we shall show that both $R_0$-non-degenerate models with $\fppo=0$ as well as all $R_0$-degenerate models feature strongly-coupled MS backgrounds with $R=R_0$.

\section{Linearised $f(R)$ EOM in a MS background}
\label{Sec3}

As is customary, we will study the linearised spectrum of both $R_0$-degenerate as well as $R_0$-non-degenerate $f(R)$ models on MS space-times by performing small perturbations around said backgrounds. To that end, we split the full metric $g_{\mu\nu}$ as 
\begin{equation}
    g_{\mu\nu}=g^\szero_{\mu\nu}+h_{\mu\nu},
\end{equation}
where $\smash{g_{\mu\nu}^\szero}$ is the MS background and $h_{\mu\nu}$ represents the perturbation. Up to linear order in $h_{\mu\nu}$, the Ricci tensor and the Ricci scalar may be expanded as follows:
\begin{eqnarray}
    R_{\mu\nu} & = & R_{\mu\nu}^\szero+R_{\mu\nu}^\Rh+\bigo(h^2), \\
    R & = &  R_0+R^\Rh+\mathcal{O}(h^2),
\end{eqnarray}
where superindex $(0)$ corresponds to quantities evaluated for the background MS metric $\smash{g_{\mu\nu}^{(0)}}$---with $R_0\equiv R^\szero$---and superindex $(h)$ corresponds to the terms linear in $h_{\mu\nu}$. These turn out to be given, respectively, by
\begin{eqnarray}
    R_{\mu\nu}^\Rh & = & -\dfrac{1}{2}\Box h_{\mu\nu} -\dfrac{1}{2} \Do_\mu \Do_\nu h +\dfrac{4\Lambda}{3} h_{\mu\nu}- \dfrac{\Lambda}{3}g_{\mu\nu}^\szero h \nonumber   \\
    & + &\Do_{(\mu|} \Do_\rho h^\rho{}_{|\nu)}
\end{eqnarray}
and
\begin{equation}
    R^\Rh=- \Box  h + \Do_\mu \Do_\nu h^{\mu\nu}-\Lambda h, 
\end{equation}
where we have introduced $\smash{h=g_\szero^{\mu\nu}h_{\mu\nu}}$ and all covariant derivatives refer to the background metric $\smash{g_{\mu\nu}^\szero}$, which is also employed to raise and lower indices. This shall be the convention to be followed hereafter.

In order to have a well-defined linear regime, and for the sake of concreteness, we shall require function $f$ to be analytical around $R_0$, i.e.
\begin{equation}
    f(R)=\fo+\fpo(R-R_0)+\ldots
\end{equation}
This allows one to expand the vacuum $f(R)$ EOM \eqref{EOM} as follows:
\begin{eqnarray} \label{eq:linearised EOM}
    0 \, & = & \fppo\left[R_{\mu\nu}^\szero-(\Do_\mu \Do_\nu-g_{\mu\nu}^\szero \Box)\right]R^\Rh \\
    & + & \fpo \left[R_{\mu\nu}^\Rh-\dfrac{R^\Rh}{2} g_{\mu\nu}^\szero\right] - \dfrac{\fo}{2} h_{\mu\nu}+\bigo(h^2), \nonumber
\end{eqnarray}
where we have taken into account that the MS background, having constant scalar curvature $R_0$, satisfies equation \eqref{eq:EOM MS}. In addition, by taking the trace of \eqref{EOM}, and expanding it at first order in $h_{\mu\nu}$, one gets:
\begin{equation} \label{trace_h}
    \fppo \left(\Box + \dfrac{R_0}{3}\right) R^\Rh-\dfrac{\fpo}{3}R^\Rh+\bigo(h^2)=0.
\end{equation}
If $\fppo\neq 0$, equation \eqref{trace_h} can be recast in the form of a canonical Klein-Gordon equation, namely
\begin{equation}
\label{Box_Rh}
(\Box - m_\eff^2) R^\Rh+\bigo(h^2)=0,
\end{equation}
where the square of the effective mass $m_\eff$ is given by
\begin{equation}
m_\eff^2=\dfrac{1}{3}\left[\dfrac{\fpo}{\fppo}-R_0\right].
\label{m2}
\end{equation}
Under these circumstances, i.e.~whenever $\fppo\neq 0$, the gauge-invariant\footnote{
    See Appendix \ref{Appendix: gauge-fixing}.
} perturbation $R^\Rh$ can be considered to be an independent scalar degree of freedom of mass $m_\eff$,\footnote{
    For a discussion on the physical interpretation of models with $m_\eff^2<0$, see Appendix \ref{Appendix: m2 < 0}.
} essentially the scalaron.\footnote{
    The relationship between the Einstein-frame scalaron and the scalar perturbation $R^\Rh$ is elucidated on Appendix \ref{appendix:scalaron}.
} Therefore, provided that $\fppo\neq 0$, both $R_0$-degenerate and $R_0$-non-degenerate $f(R)$ models can in principle propagate a scalar degree of freedom on top of MS backgrounds with $R=R_0$. Otherwise, if $\fppo\neq 0$, MS backgrounds with $R=R_0$ exhibit a strong-coupling instability in both $R_0$-degenerate and $R_0$-non-degenerate $f(R)$ models.

Up to this point, every computation we have performed is fully general and valid for any $f(R)$ model, be it $R_0$-degenerate or $R_0$-non-degenerate. However, it is now convenient to specialise for each type of model separately so as to better appreciate the effect of degeneration in the propagation of linear degrees of freedom.

\section{Non-degenerate $f(R)$ models}
\label{sec:non-degenerate}

In the case of $R_0$-non-degenerate $f(R)$ models, i.e.~$\fpo\neq 0$, the linearised EOM \eqref{eq:linearised EOM} simplify considerably if one introduces the $\bigo(h)$ quantity
\begin{equation} 
\label{eq:barh}
    \barh_{\mu\nu} \equiv h_{\mu\nu} - \left[\dfrac{h}{2} + \dfrac{\fppo}{\fpo} R^\Rh\right] g^\szero_{\mu\nu},
\end{equation}
which differs from the usual trace-reversed perturbation employed in GR by the term proportional to $R^\Rh$; in fact, one may immediately check that $\bar h \neq - h$. We stress that \eqref{eq:barh} is only well-defined for non-degenerate models, where it is possible to divide by $\fpo\neq 0$.\footnote{
    Alternatively, one could have defined $\barh_{\mu\nu}$ as
    \begin{equation*}
        \barh_{\mu\nu} \equiv \fpo\left(h_{\mu\nu} - \dfrac{h}{2}g^\szero_{\mu\nu}\right) - \fppo R^\Rh g^\szero_{\mu\nu},
    \end{equation*}
    as done in \cite{Casado-Turrion:2023rni}. Unlike \eqref{eq:barh}, this expression for $\barh_{\mu\nu}$ is valid for both degenerate and non-degenerate $f(R)$ models.
}

After some algebra, it is possible to find that, for non-degenerate models, first-order EOM \eqref{eq:linearised EOM} can be rewritten in terms of $\barh_{\mu\nu}$ as
\begin{eqnarray} \label{eq:ND linearised EOM}
    0 \, & = & \left(\Box-\dfrac{R_0}{6}\right)\barh_{\mu\nu}+\dfrac{R_0}{6}\barh g ^\szero_{\mu\nu}-2\Do_{(\mu|} \Do_\rho\, \barh^\rho{}_{|\nu)} \nonumber \\
    & + & g ^\szero_{\mu\nu}\Do_\rho \Do_\sigma\, \barh^{\rho\sigma} + \bigo(h^2),
\end{eqnarray}
where have made use of \eqref{eq:EOM trace MS}.

As shown in Appendix \ref{Appendix: gauge-fixing}, it is always possible to perform gauge transformations such that the ensuing conditions are all simultaneously satisfied:
\begin{equation} \label{eq:TT + h_mu0}
    \Do^\mu\barh_{\mu\nu}=0, \mybigskip\barh=0, \mybigskip\barh_{\mu0}=0.
\end{equation}
In such scenario, linear EOM \eqref{eq:ND linearised EOM} simplify considerably, reducing to
\begin{equation} \label{EOM_h_tilde}
    \left(\Box - \dfrac{R_0}{6}\right)\barh_ {\mu\nu}+\bigo(h^2)=0, 
\end{equation}
which describe the propagation of an apparently massive graviton $\barh_{\mu\nu}$. Nonetheless, as also discussed in Appendix \ref{Appendix: gauge-fixing}, conditions \eqref{eq:TT + h_mu0} ensure that $\barh_{\mu\nu}$ contains only two gauge-independent components, corresponding to the standard massless and traceless graviton also found in GR. More precisely, using the gauge-fixing conditions \eqref{eq:TT + h_mu0} above, the only non-vanishing components of the linearised  metric fluctuations $\barh_{\mu\nu}$ are the purely spatial ones $\barh_{ij}$, subject to the constraints
\begin{equation} \label{TT conditions dS}
    \delta^{ij}h_{ij}=0,\mybigskip\Do^i h_{ij}=0,
\end{equation}
thus leaving two physical degrees of freedom associated to the graviton, as mentioned before. Once the previous constraints have been taken into account, the only remaining non-trivial components of equation \eqref{EOM_h_tilde} for the two graviton degrees of freedom are
\begin{equation} \label{Box_h_tilde_ij}
    \left(\Box - \dfrac{R_0}{6}\right)\barh_{ij}+\bigo(h^2)=0.
\end{equation}
In summary, we have the following:

\begin{result} \label{result:DOF number ND}
    The total number of independent, gauge-invariant degrees of freedom propagating on top of MS backgrounds with $R=R_0$ in $R_0$-non-degenerate $f(R)$ models is either three (graviton $+$ scalaron), provided that $\fppo\neq 0$, or only two (the graviton), if $\fppo=0$.
\end{result}

\subsection{Scalaron and graviton equations on a dS background space-time in planar coordinates} \label{sec:scalaron and graviton eqs in dS}

In order to better illustrate the physical significance of these results, let us now concentrate---without any loss of  generality---on the case of a dS background ($R_0>0$), whose line element expressed in so-called planar coordinates $x^\mu=(t,\vec{x})$ reads
\begin{equation} \label{dS}
    \dif s^2_\szero=-\dif t^2+a^2(t)\,\dif\vec{x}^{2},
\end{equation}
where
\begin{equation} \label{a_dS}
    a(t)=\e^{H_0 t},
\end{equation}
with $H_0$ being the usual dS Hubble constant, i.e.
\begin{equation} \label{eq:H_0}
    H_0\equiv\sqrt{\dfrac{\Lambda}{3}}=\sqrt{\dfrac{R_0}{12}}.
\end{equation}
Recall that, in non-degenerate models, $R_0$---and thus $\Lambda$ and $H_0$---are related to $\fo$ and $\fpo$ through \eqref{eqn:Lambda ND}.

It is straightforward to show that, when using planar coordinates $(t,\vec{x})$, Klein-Gordon equation \eqref{Box_Rh} for the scalaron becomes
\begin{equation} \label{dS scalar eqn ND}
    \left(-\partial_t^2 + \dfrac{\nabla^{2}}{a^2(t)}- 3 H_0 \partial_t-m_\eff^2\right)R^\Rh+\bigo(h^2)=0.
\end{equation}
However, the graviton equations \eqref{Box_h_tilde_ij} do not have such a simple form in planar coordinates, since the action of d'Alembert's operator on a tensor such as $\barh_{ij}$ is highly non-trivial, leading to convoluted expressions. This issue can be nevertheless remedied by means of the following field redefinition \cite{Dodelson:2003ft,Yang:2011cp}:
\begin{equation} \label{eq: H_ij}
    \barh_{ij}\equiv a^2(t) H_{ij}.
\end{equation}
It is then possible to check that, in terms of $H_{ij}$ and using planar coordinates $(t,\vec{x})$, EOM \eqref{Box_h_tilde_ij} become
\begin{equation} \label{tensor_dalembert}
    \left(-\partial_t^2 + \dfrac{\nabla^{2}}{a^2(t)}- 3 H_0 \partial_t\right) H_{ij}+\bigo(h^2)=0,
\end{equation}
which is reminiscent to equation \eqref{dS scalar eqn ND} for the scalar degree of freedom, albeit with vanishing mass.

As discussed in Appendix \ref{Appendix: mode solutions}, simple mode solutions of equations of the form \eqref{dS scalar eqn ND} or \eqref{tensor_dalembert} with fixed wave vector may be readily obtained, whereby wave-packets representing the actual, localised perturbations can be constructed through superposition.\footnote{
    We remind the reader superposing mode solutions is only possible because we are working on the linear approximation to the full theory.
} Equipped with the results therein, we will now discuss the mode solutions to equations \eqref{dS scalar eqn ND} and \eqref{tensor_dalembert} and analyse the stability of the graviton and scalaron degrees of freedom.

The busy reader who would prefer to omit the ensuing computational details is invited to proceed directly to Results \ref{result:graviton stability ND} and \ref{result:scalaron stability ND}.

\subsection{Mode solutions to the linearised graviton EOM and their stability}
\label{sec:graviton ND}

As mentioned before, equation \eqref{tensor_dalembert} for the re-scaled graviton modes $H_{ij}$---as defined in \eqref{eq: H_ij}---corresponds to the particular case of equation \eqref{eq:KG in dS} in Appendix \ref{Appendix: mode solutions} where $m=0$. Hence, using solutions \eqref{eq:zero-mode} and \eqref{eq:k-mode} for \eqref{eq:KG in dS} (particularised for $m=0$), as well as equation \eqref{eq: H_ij}, the decomposition of the physical graviton field $\barh_{ij}$ into modes \smash{$\barh_{ij}^{\vec{k}}$} with well-defined wave vector \smash{$\vec{k}$} is given by
\begin{eqnarray} \label{eq:ND graviton modes}
    \barh_{ij}^{\vec{k}}(t,\vec{x}) & = & A_{ij}^\sone(\vec{k})\,\e^{H_0 t/2}\,\psi^\sone_{\vec{k}}(t)\,\e^{-\iu\vec{k}\cdot\vec{x}} \nonumber \\
    & + & A_{ij}^\stwo(\vec{k})\,\e^{H_0 t/2}\,\psi^\stwo_{\vec{k}}(t)\,\e^{+\iu\vec{k}\cdot\vec{x}},
\end{eqnarray}
where, in the expression above,
\begin{eqnarray}
    \psi^\sboth_{\vec{k}=\vec{0}}(t) & = & \e^{\mp\Omega_0 t}, \label{eq:ND graviton psi12 k = 0} \\
    \psi^\sboth_{\vec{k}\neq\vec{0}}(t) & = & \Hboth_{3/2}\Left(\dfrac{|\vec{k}|}{H_0}\,\e^{-H_0 t}\Right), \label{eq:ND graviton psi12 k neq 0}
\end{eqnarray}
with $\Omega_0$ being given by \eqref{eq:Omega_0} and $\Hboth_\nu$ respectively being the Hankel functions of the first and second kind with index $3/2$.\footnote{
    Damping factor $\Omega_0$ in \eqref{eq:ND graviton psi12 k = 0} results from setting $m=0$ in \eqref{eq:omega_0}. Similarly, the particular value $3/2$ for the index in \eqref{eq:ND graviton psi12 k neq 0} arises from evaluating \eqref{eq:nu2} for $m=0$.
} Additionally, tensors $A_{ij}^\sboth(\vec{k})$ appearing in \eqref{eq:ND graviton modes} encode the amplitude of each mode, and must satisfy
\begin{equation} \label{eq:ND graviton A_ij constraints 1}
    \delta^{ij}A_{ij}^\sboth(\vec{k})=0,\mybigskip k^i A_{ij}^\sboth(\vec{k})=0
\end{equation}
in order to fulfil TT conditions \eqref{TT conditions dS}, as well as symmetry constraint
\begin{equation} \label{eq:ND graviton A_ij constraints 2}
    A_{ji}^\sboth(\vec{k})=A_{ij}^\sboth(\vec{k}),
\end{equation}
and the various case-dependent conditions guaranteeing the reality of modes \smash{$\barh_{ij}^{\vec{k}}$}, which are discussed in full detail in Appendix \ref{Appendix: mode solutions}. As such, a wave propagating on the \smash{$\vec{k}$} direction encapsulates two transverse degrees of freedom, conventionally denoted $A_+$ and $A_\times$, which correspond to the two standard polarisations of massless and traceless spin-2 gravitons, as stated before.

The stability analysis of graviton modes \eqref{eq:ND graviton modes} proceeds as follows. First, the zero-mode \smash{$\barh_{ij}^{\vec{0}}$} is always tachyonic, having a component---namely, the second one---which grows exponentially as time progresses, as can be clearly seen by expressing the mode solely in terms of $H_0$ using \eqref{eq:Omega_0}:
\begin{equation} \label{eq:graviton zero-mode tachyonic}
    \barh_{ij}^{\vec{0}}(t) = A_{ij}^\sone(\vec{0})\,\e^{-H_0 t} + A_{ij}^\stwo(\vec{0})\,\e^{2H_0 t}.
\end{equation}
However, as $t\rightarrow\infty$, the graviton zero-mode merely grows as $a^2(t)=\e^{2H_0 t}$, which is precisely the same rate at which the background dS space-time \eqref{dS} expands, i.e.
\begin{equation}
    g_{ij}^\szero=\e^{2H_0 t}\,\delta_{ij}.
\end{equation}
Because the zero-mode does \emph{not} grow faster than the background, condition \smash{$|h_{ij}^{\vec{0}}|/|g_{ij}^\szero|\ll 1$} is satisfied as time progresses, and it can be therefore concluded that there is no future tachyonic instability; the accelerated expansion of the background dS space-time dilutes the tachyonic growth of zero-mode perturbations.

At this point, a crucial observation must be made, which will be relevant for the remainder of our stability analysis, both in the graviton and scalaron cases. As is evident from \eqref{eq:graviton zero-mode tachyonic}, the first component of the zero-mode diverges in the remote past ($t\rightarrow-\infty$). However, this exponential growth---akin to a tachyonic instability---should be disregarded as unphysical, since we intend to consider only perturbations produced at some finite $t=t_0$ and propagating from that instant onwards. Therefore, we will henceforth not consider modes which are unbounded in the far past as unstable, unless they also grow faster than the background as time advances.

Finally, in order to assess the stability of graviton modes with \smash{$\vec{k}\neq\vec{0}$}, it is instrumental to note that Hankel functions \smash{$\Hboth_{n+1/2}$} (where $n$ is an integer) can be expressed in terms of elementary functions. In particular, for real $z$, one has
\begin{equation}
    \Hone_{3/2}(z) = \left[\Hone_{3/2}(z)\right]^* = - \sqrt{\dfrac{2}{\pi z}}\,\e^{+\iu z}\left(1+\dfrac{\iu}{z}\right).
\end{equation}
Therefore, performing a coordinate transformation from `planar time' $t$ to so-called `conformal time' $\eta$, defined as
\begin{equation} \label{eq:eta}
    \eta\equiv\dfrac{\e^{-H_0 t}}{H_0},
\end{equation}
setting \smash{$z=|\vec{k}|\eta$}, and suitably redefining \smash{$A^\sboth_{ij}(\vec{k})$} so that all numerical pre-factors are absorbed into the amplitude tensors, we find that the graviton \smash{$\vec{k}$}-modes are given by
\begin{eqnarray}
    \barh_{ij}^{\vec{k}}(\eta,\vec{x}) & = & A_{ij}^\sone(\vec{k})\,\dfrac{|\vec{k}|\eta+\iu}{(|\vec{k}|\eta)^2}\,\e^{+\iu(|\vec{k}|\eta-\vec{k}\cdot\vec{x})} \nonumber \\
    & + & A_{ij}^\stwo(\vec{k})\,\dfrac{|\vec{k}|\eta-\iu}{(|\vec{k}|\eta)^2}\,\e^{-\iu(|\vec{k}|\eta+\vec{k}\cdot\vec{x})}.
\end{eqnarray}
These modes correspond to damped plane waves whose amplitude ($i$) vanishes when \smash{$\eta\rightarrow+\infty$}, i.e.~in the far past $t\rightarrow-\infty$, and ($ii$) diverges as \smash{$\eta^{-2}\propto a^2(\eta)$} when \smash{$\eta\rightarrow+\infty$}, i.e.~in the distant future $t\rightarrow+\infty$. Similarly to the zero-wave-vector case, we find that graviton \smash{$\vec{k}$}-modes do not grow faster than the dS background space-time \eqref{dS}, whose line element in coordinates \smash{$(\eta,\vec{x})$} is given by
\begin{equation}
    \dif s^2_\szero=\left(H_0 \eta\right)^{-2}(-\dif\eta^2+\dif\vec{x}^{2}).
\end{equation}
In consequence, we have the following:

\begin{result} \label{result:graviton stability ND}
    All graviton modes \eqref{eq:ND graviton modes} propagate stably on a dS background space-time with $R=R_0$ in $R_0$-non-degenerate $f(R)$ models, regardless of whether \smash{$\vec{k}=\vec{0}$} or \smash{$\vec{k}\neq\vec{0}$}.
\end{result}

\subsection{Mode solutions to the linearised scalaron EOM and their stability}
\label{sec:scalaron ND}

Having already considered the two spin-2 degrees of freedom enclosed in the metric perturbation, we turn to investigate the scalaron fluctuation related to $R^\Rh$ which propagates in non-degenerate $f(R)$ models with $\fppo\neq 0$. As previously stated in Subsection \ref{sec:scalaron and graviton eqs in dS}, EOM \eqref{dS scalar eqn ND} for the scalar degree of freedom is a particular case of equation \eqref{eq:KG in dS} in Appendix \ref{Appendix: mode solutions}, with $m=m_\eff$ as given by \eqref{m2}. Owing to this, simple mode solutions \smash{$R^\Rh_{\vec{q}}$} to \eqref{dS scalar eqn ND} with wave vector \smash{$\vec{q}$} can be easily found using, once again, results \eqref{eq:zero-mode} and \eqref{eq:k-mode} encapsulated in Appendix \ref{Appendix: mode solutions}. In particular, we find:
\begin{eqnarray} \label{eq:ND scalaron modes}
    R^\Rh_{\vec{q}}(t,\vec{x}) & = & A^\sone(\vec{q})\,\e^{-3H_0 t/2}\,\psi^\sone_{\vec{q}}(t)\,\e^{-\iu\vec{q}\cdot\vec{x}} \nonumber \\
    & + & A^\stwo(\vec{q})\,\e^{-3H_0 t/2}\,\psi^\stwo_{\vec{q}}(t)\,\e^{+\iu\vec{q}\cdot\vec{x}},
\end{eqnarray}
with
\begin{eqnarray}
    \psi^\sboth_{\vec{q}=\vec{0}}(t) & = & \e^{\pm\iu\omega_0 t}, \label{eq:ND scalaron psi12 k = 0} \\
    \psi^\sboth_{\vec{q}\neq\vec{0}}(t) & = & \Hboth_\nu\Left(\dfrac{|\vec{q}|}{H_0}\,\e^{-H_0 t}\Right), \label{eq:ND scalaron psi12 k neq 0}
\end{eqnarray}
where $\omega_0$ and $\nu$ are respectively given by \eqref{eq:omega_0}--\eqref{eq:Omega_0} and \eqref{eq:nu2} (particularised for $m=m_\eff$). Mode amplitudes \smash{$A^\sboth(\vec{q})$} must also fulfil the case-dependent conditions assuring that \smash{$R^\Rh_{\vec{q}}$} is real; said conditions are discussed exhaustively throughout Appendix \ref{Appendix: mode solutions}.

As done in the graviton case, we start our stability analysis of the scalar mode with the zero-mode. Two scenarios must be distinguished within this case. On the one hand, if $\omega_0^2<0$, corresponding---as per \eqref{m2}, \eqref{eq:H_0}, \eqref{eq:omega_0} and \eqref{eq:Omega_0}---to
\begin{equation} \label{ND scalaron omega_0^2 < 0}
    R_0 > \dfrac{16\fpo}{25\fppo},
\end{equation}
the exponentials in \eqref{eq:ND scalaron psi12 k = 0} become real, and scalaron modes \eqref{eq:ND scalaron modes} are given by
\begin{eqnarray} \label{eq:scalaron zero-mode tachyonic}
    \Left.R^\Rh_{\vec{0}}(t,\vec{x})\Right|_{\omega_0^2<0} & = & A^\sone(\vec{0})\,\e^{-3H_0 t/2}\,\e^{-|\omega_0| t} \nonumber \\
    & + & A^\stwo(\vec{0})\,\e^{-3H_0 t/2}\,\e^{+|\omega_0|t}.
\end{eqnarray}
The first component in \eqref{eq:scalaron zero-mode tachyonic} decays exponentially in time and is thus stable. However, the second component grows unboundedly as $t\rightarrow+\infty$ unless
\begin{equation}
    -\dfrac{3H_0}{2}+|\omega_0|=-\Omega_0+\sqrt{\Omega_0^2-m_\eff^2}\leq 0.
\end{equation}
Since $\Omega_0$ is positive, the inequality above is only satisfied provided that
\begin{equation}
    m_\eff^2\geq 0\myskip\Longleftrightarrow\myskip R_0\leq\dfrac{\fpo}{\fppo}.
\end{equation}
On the other hand, if $\omega_0^2\geq 0$ (which is only possible if $m_\eff^2\geq\Omega_0^2>0$), it is clear from \eqref{eq:ND scalaron modes} and \eqref{eq:ND scalaron psi12 k = 0} that the zero-mode \smash{$R^\Rh_{\vec{0}}|_{\omega_0^2\geq 0}$} consists of damped plane waves whose frequency $\omega_0$ decays as $t\rightarrow+\infty$. For this reason, we conclude that the scalaron mode with vanishing wave vector is stable provided that the scalaron mass squared $m_\eff^2$ is non-negative.

Two comments on the previous result are in place. First, notice that it is necessary to require the scalar modes \smash{$R^\Rh_{\vec{q}}$} to be strictly constant or decreasing in time in order to avoid a tachyonic instability. This is because, for the scalaron, the background is constant (namely, $R^\szero=R_0$), in contrast with the graviton case, whose corresponding background \smash{$g_{\mu\nu}^\szero$} expands exponentially in time (entailing that the metric perturbation could be considered stable as long as its growth rate is slower than that of the background dS space-time). Second, and in analogy with the graviton case, the scalaron zero-mode blows up as $t\rightarrow-\infty$, and this will turn out to be the case regardless of the values of \smash{$\vec{q}$}, $m_\eff^2$ and $\omega_0^2$. This issue should be ignored on physical grounds for the reasons stated above in Subsection \ref{sec:graviton ND}.

Regarding the scalaron modes with \smash{$\vec{q}\neq\vec{0}$}, it is once again convenient to express them using conformal time $\eta$---as defined in \eqref{eq:eta}---rather than planar time $t$:
\begin{eqnarray} \label{eq:ND scalaron modes q neq 0 conformal}
    R^\Rh_{\vec{q}}(\eta,\vec{x}) & = & A^\sone(\vec{q})\,(H_0\eta)^{3/2}\,\Hone_{\nu}(|\vec{q}|\eta)\,\e^{-\iu\vec{q}\cdot\vec{x}} \nonumber \\
    & + & A^\stwo(\vec{q})\,(H_0\eta)^{3/2}\,\Htwo_{\nu}(|\vec{q}|\eta)\,\e^{+\iu\vec{q}\cdot\vec{x}}.
\end{eqnarray}
The stability of the scalaron \smash{$\vec{q}$}-modes can be determined using the expansion of Hankel functions for small values of their argument, i.e.~$0<|\vec{q}|\eta\ll 1$.\footnote{
    The opposite limit, i.e.~$|\vec{q}|\eta\gg 1$, corresponding to the far-past behaviour of the scalar \smash{$\vec{q}$}-modes, reveals once again an unphysical divergence thereof as $t\rightarrow-\infty$.
} The precise form of said expansion depends on whether index $\nu$ is either vanishing (equivalent to $\omega_0^2=0$), positive (entailing that $\omega_0^2<0$) or pure imaginary (corresponding to $\omega_0^2>0$). Therefore, we shall contemplate the three aforementioned scenarios separately.

First, for $\nu=0$, the behaviour of Hankel functions for small values of their argument is
\begin{equation}
    \Hboth_0(z)\underset{0<z\ll 1}{\sim} 1 \pm \dfrac{2\iu}{\pi}\left[\gamma+\ln\left(\dfrac{z}{2}\right)\right],
\end{equation}
where $\gamma$ is the Euler-Mascheroni constant. As a result, the \smash{$\vec{q}$}-modes behave schematically as
\begin{eqnarray}
    \Left.R^\Rh_{\vec{q}}\Right|_{\nu=0} & \overset{0<|\vec{q}|\eta\ll 1}{\sim} & (|\vec{q}|\eta)^{3/2}+(|\vec{q}|\eta)^{3/2}\,\ln(|\vec{q}|\eta) \nonumber \\
    & \underset{|\vec{q}|\eta\rightarrow 0}{\longrightarrow} & 0,
\end{eqnarray}
and thus the scalaron modes with \smash{$\vec{q}\neq\vec{0}$} are stable when $\nu=0$. Observe that, in this case, $m_\eff^2=\Omega_0^2>0$.

Next, for $\nu>0$, both Hankel functions $\Hboth_\nu$ admit an expansion of the form
\begin{equation}
    \Hboth_{\nu>0}(z)\underset{0<z\ll 1}{\sim} C^\sboth_+(\nu)\,z^{+\nu}+C^\sboth_-(\nu)\,z^{-\nu},
\end{equation}
where \smash{$C^\sboth_\pm(\nu)$} are numerical coefficients depending on the particular value of $\nu$. Because of this, the asymptotic behaviour of the modes is given schematically by
\begin{eqnarray}
    \Left.R^\Rh_{\vec{q}}\Right|_{\nu>0} & \overset{0<|\vec{q}|\eta\ll 1}{\sim} & C_+(\nu)\,(|\vec{q}|\eta)^{3/2+\nu} \nonumber \\
    & \quad\quad + & C_-(\nu)\,(|\vec{q}|\eta)^{3/2-\nu}.
\end{eqnarray}
The terms proportional to \smash{$(|\vec{q}|\eta)^{3/2+\nu}$} all vanish in the limit \smash{$|\vec{q}|\eta \rightarrow 0$} for every positive $\nu$, but the terms proportional to \smash{$(|\vec{q}|\eta)^{3/2-\nu}$} only tend to zero as \smash{$|\vec{q}|\eta \rightarrow 0$} for $\nu\leq 3/2$, corresponding to
\begin{equation}
    \omega_0^2\geq-\Omega_0^2 \myskip\Longleftrightarrow\myskip m_\eff^2\geq 0.
\end{equation}
Therefore, the scalaron \smash{$\vec{q}$}-modes are stable in the case $\nu>0$ if and only if $m_\eff^2$ is non-negative, in consonance with our previous findings.

Finally, for pure-imaginary $\nu$ (i.e.~$\nu=\iu|\nu|$), we have that the expansions for the Hankel functions and the scalaron \smash{$\vec{q}$}-modes are very similar to those in the last scenario. Schematically,
\begin{eqnarray}
    \Left.R^\Rh_{\vec{q}}\Right|_{\nu=\iu|\nu|} & \overset{0<|\vec{q}|\eta\ll 1}{\sim} & C_+(\nu)\,(|\vec{q}|\eta)^{3/2+\iu|\nu|} \nonumber \\
    & \quad\quad + & C_-(\nu)\,(|\vec{q}|\eta)^{3/2-\iu|\nu|},
\end{eqnarray}
regardless of the precise determination chosen to assign a unique value to multivaluate complex powers \smash{$(|\vec{q}|\eta)^{\pm\iu|\nu|}$}. By recasting, without loss of generality,
\begin{equation}
    (|\vec{q}|\eta)^{\pm\iu|\nu|} = \cos\Left[|\nu|\ln\left(|\vec{q}|\eta\right)\Right] \pm \iu\sin\Left[|\nu|\ln\left(|\vec{q}|\eta\right)\Right],
\end{equation}
we have that each of the \smash{$\vec{q}$}-modes represents a wave of decreasing amplitude and increasing frequency as \smash{$|\vec{q}|\eta\rightarrow 0$}. Given that scalaron modes with \smash{$\vec{q}\neq\vec{0}$} vanish in the far-future limit, we conclude that they are stable for $\nu=\iu|\nu|$. Because this case corresponds to $m_\eff^2>\Omega_0^2>0$, we find one more time that the scalaron is stable for $m_\eff^2>0$. This completes the stability assessment for the scalaron modes with non-vanishing wave vector, which have turned out to be stable as long as the scalaron mass is non-negative.

To summarise the findings in this Section, we have found the following Result:

\begin{result} \label{result:scalaron stability ND}
    All scalaron modes \eqref{eq:ND scalaron modes} propagating on top of a dS background space-time with $R=R_0$ are stable in $R_0$-non-degenerate $f(R)$ models such that $m_\eff^2$---as given by \eqref{m2}---is non-negative; this condition is tantamount to
    \begin{equation} \label{ND scalaron m^2 >= 0}
        R_0\leq\dfrac{\fpo}{\fppo}.
    \end{equation}
    If $m_\eff^2<0$, modes \eqref{eq:ND scalaron modes} are tachyonically unstable.
\end{result}

\begin{figure*}[t]
    \centering
    \includegraphics[width=0.875\linewidth]{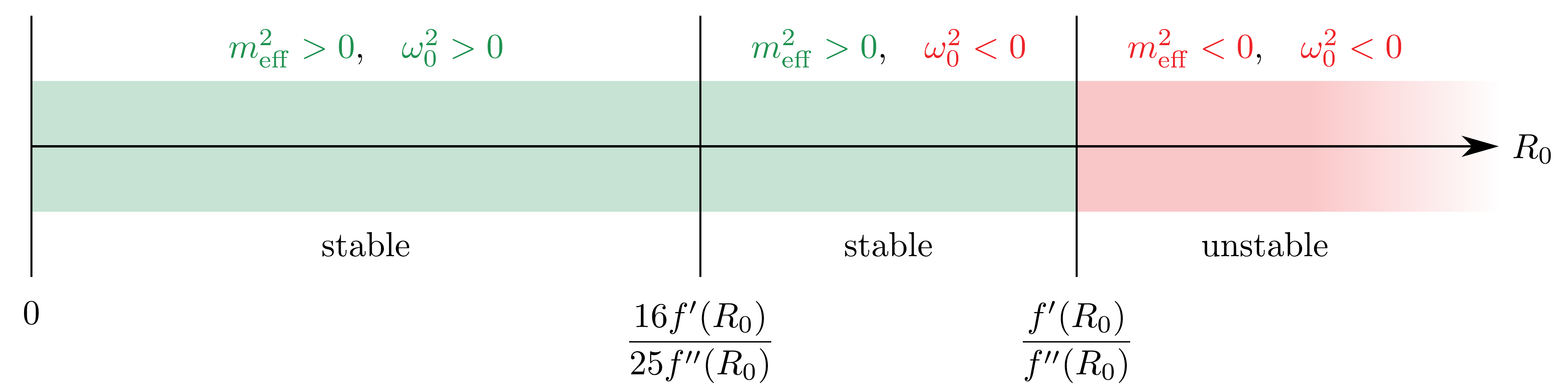}
    \caption{Scalaron stability in terms of the dS background curvature, $R_0$, in non-degenerate $f(R)$ models with $\fppo\neq 0$. Notice that, as discussed in the text, there are some values of $R_0$ such that $\omega_0^2<0$ but $m_\eff^2>0$, in which case the cosmic expansion compensates the tachyonic character of perturbations.}
    \label{fig:scalaron stability}
\end{figure*}

Our Result \ref{result:scalaron stability ND} thus evokes its counterpart in Minkowski, namely, that the avoidance of tachyonic instabilities in the theory of a massive scalar field in flat space-time is guaranteed by the non-negativity of the field's mass squared. Notice moreover that, contrary to the situation in Minkowski space-time, it is possible to have tachyonic scalar modes which are stable on a dS background. This is because for background curvatures $R_0$ satisfying
\begin{equation}
    \dfrac{16\fpo}{25\fppo} < R_0 \leq \dfrac{\fpo}{\fppo},
\end{equation}
we have that $\omega_0^2<0$ but $m_\eff^2\geq 0$, as per \eqref{ND scalaron omega_0^2 < 0} and \eqref{ND scalaron m^2 >= 0}; a pictorial representation of this fact is provided in Figure \ref{fig:scalaron stability}. Similarly to the graviton case, this phenomenon can be entirely attributed to the accelerated expansion of the dS background space-time, which might be strong enough to compensate the exponential growth of tachyonic modes, depending on the value of the scalaron mass. For further insight on the interpretation of $f(R)$ models such that $m_\eff^2<0$ in terms of the Einstein-frame representation, we refer the reader to Appendix \ref{Appendix: m2 < 0}.

\section{Degenerate $f(R)$ models}
\label{Sec5}

Having analysed MS backgrounds in non-degenerate $f(R)$ models thoroughly, we shall now proceed to investigate the degenerate case, characterised---as per \eqref{eq:EOM trace MS}---by $\fo=0$ and $\fpo=0$ for some $R=R_0$. Under these conditions, linearised EOM \eqref{eq:linearised EOM} become
\begin{equation} \label{eq:linearised EOM D}
    \fppo\left[R_{\mu\nu}^\szero-(\Do_\mu \Do_\nu-g_{\mu\nu}^\szero \Box)\right]R^\Rh+\bigo(h^2)=0,
\end{equation}
while their trace \eqref{trace_h} turns into
\begin{equation} \label{eq:linearised EOM trace D}
    \fppo \left(\Box + \dfrac{R_0}{3}\right) R^\Rh+\bigo(h^2)=0.
\end{equation}
We readily notice that all terms depending on $h_{\mu\nu}$ have disappeared from \eqref{eq:linearised EOM D} and \eqref{eq:linearised EOM trace D}; only \smash{$R^\Rh$} remains therein. In consequence, MS backgrounds with $R=R_0$ are strongly-coupled in all $R_0$-degenerate $f(R)$ models, given that their linearised spectrum lacks the massless and traceless graviton appearing in non-degenerate $f(R)$ models with $\fppo\neq 0$. This is the generalisation of the results first established in \cite{Casado-Turrion:2023rni} to those scenarios where $R_0\neq 0$. Another immediate consequence of EOM \eqref{eq:linearised EOM D} and \eqref{eq:linearised EOM trace D} is that the scalaron also disappears from the linear spectrum of the model if $\fppo=0$, in complete analogy with the non-degenerate case.

Specialising now to $R_0$-degenerate models such that $\fppo\neq 0$, we find that the scalaron $R^\Rh$ satisfies Klein-Gordon equation \eqref{eq:linearised EOM trace D},\footnote{
    Notice that the effective scalaron mass squared is $m_\eff^2=-R_0/3$ for $R_0$-degenerate models.
} alongside a set of constraints provided by \eqref{eq:linearised EOM D}, which constitute the degenerate-model counterpart of the graviton EOM \eqref{eq:ND linearised EOM} present in non-degenerate models. As such, even though one could in principle think that $R_0$-degenerate $f(R)$ models with $\fppo\neq 0$ propagate just a single degree of freedom, it turns out that this is not the case due to the presence of constraints \eqref{eq:linearised EOM D}, as we will see now.

\subsection{$(R_0=0)$-degenerate models}
\label{Sec5a}

Let us start by considering the simple case $R_0=0$, where the background space-time \smash{$g^\szero_{\mu\nu}$} reduces to the Minkowski metric $\eta_{\mu\nu}$. Then, we can choose Cartesian-like coordinates $x^\mu=\left(t, \vec x\right)$ whereby EOM \eqref{eq:linearised EOM trace D} for $R^\Rh$ becomes, assuming $f''(0)\neq 0$,
\begin{equation} \label{eq:D EOM Minkowski}
    \Box R^\Rh+\bigo(h^2)=0,
\end{equation}
while constraints \eqref{eq:linearised EOM D} transform into
\begin{equation}
    (\partial_\mu \partial_\nu -\eta_{\mu\nu}\Box) R^\Rh+\bigo(h^2)=0.
\end{equation}
The constraints simplify even further upon substitution of EOM \eqref{eq:D EOM Minkowski}:
\begin{equation} \label{eq:D constraints Minkowski}
    \partial_\mu \partial_\nu R^\Rh+\bigo(h^2)=0.
\end{equation}
Equations \eqref{eq:D EOM Minkowski} and \eqref{eq:D constraints Minkowski} can be integrated immediately to find the general solution
\begin{equation} \label{scalaron D Minkowski}
    R^\Rh= C_\mu  x^\mu + D.
\end{equation}
Clearly, this solution is not bounded, and therefore cannot represent a localised perturbation propagating on top of the Minkowski background unless $C_\mu=0$ and $D=0$. In other words, the only linear-level solution for the scalaron fulfilling adequate boundary conditions at infinity is $R^\Rh=0$. For this reason, the following Result holds:

\begin{result} \label{result:R_0 = 0 D stability}
    The linearised spectrum of $(R_0=0)$-degenerate $f(R)$ models around Minkowski space-time is empty, since (i) they always feature a strongly-coupled graviton, (ii) if $f''(0)=0$, the scalaron is also strongly-coupled, and (iii) in cases where $f''(0)\neq 0$, the only admissible solution for the scalaron perturbation is the vanishing one.\footnote{
        Given that the disappearance of the scalaron in $(R_0=0)$-degenerate models with $f''(0)\neq 0$ is entirely due to the fact that the only possible solution for $R^\Rh$ is the identically-null function, it is debatable whether one might consider the scalaron to be strongly-coupled in such scenarios, at least from a purely terminological point of view (the graviton, nonetheless, is still strongly coupled in these models).
    }
\end{result}

Therefore, apart from being unstable, $(R_0=0)$-degenerate models are incompatible with gravitational-wave observations. Moreover, we note that all previous results concerning the lack of propagating degrees of freedom atop a Minkowski background in the purely-quadratic $f(R)$ model $f(R)=\alpha R^2$ are a particular instance of the much more general Result \ref{result:R_0 = 0 D stability} above, because $f(R)=\alpha R^2$ is an $(R_0=0)$-degenerate model.

\subsection{$(R_0\neq 0)$-degenerate models}
\label{Sec5b}

We now turn to investigate the $(R_0\neq 0)$-degenerate case with $\fppo\neq 0$. Without loss of generality, we particularise for $R_0>0$, i.e.~for a dS background. In such scenario, EOM \eqref{eq:linearised EOM trace D} can be expressed in terms of the Hubble constant $H_0$---as given by \eqref{eq:H_0}---and planar coordinates $(t,\vec{x})$ as
\begin{equation} \label{eq:D EOM dS}
    \left(-\partial_t^2 + \dfrac{\nabla^{2}}{a^2(t)}- 3 H_0 \partial_t+4H_0^2\right)R^\Rh+\bigo(h^2)=0.
\end{equation}
Additionally, constraints \eqref{eq:linearised EOM D} become
\begin{equation}
\left[\Do_\mu \Do_\nu +H^2_0\,g^\szero_{\mu\nu}\right] R^\Rh +\bigo(h^2) = 0,
\end{equation}
which can be split into three distinct equations using coordinates $(t,\vec{x})$:
\begin{eqnarray} \label{eq:D constraints dS}
0 & = & (\partial^2_t-H_0^2)  R^\Rh+\bigo(h^2),    \nonumber \\
0 & = & \partial_i(\partial_t+H_0)  R^\Rh+\bigo(h^2),    \\
0 & = & \Left[\partial_i\partial_j-\delta_{ij} a^2(t) H_0 (\partial_t+H_0)\Right]R^\Rh+\bigo(h^2). \myskip \nonumber
\end{eqnarray}
It is not difficult to find the general solution to the system formed by equations \eqref{eq:D EOM dS} and \eqref{eq:D constraints dS} above, which is
\begin{equation} \label{scalaron D dS}
    R^\Rh  = \mathcal{A}\,\e^{H_0t},
\end{equation}
where $\mathcal{A}$ is a real integration constant representing the solution's amplitude. This resembles the non-degenerate-case mode solutions in \eqref{eq:ND scalaron modes}--\eqref{eq:ND scalaron psi12 k = 0} for the particular case \smash{$\vec{q}=\vec{0}$}. However, it must be pointed out that there is a crucial difference between the solutions in \eqref{eq:ND scalaron modes}--\eqref{eq:ND scalaron psi12 k = 0} and \eqref{scalaron D dS}, namely, that the former are mode solutions to the non-degenerate scalaron EOM (which can be combined to form localised wave-packets, or to express the general solution as an infinite superposition of modes), whereas the latter represents the full, general solution for the degenerate-case scalaron EOM, similarly to \eqref{scalaron D Minkowski}. As such, the presence of an exponential in expression \eqref{scalaron D dS} does not signal the existence of a tachyonic instability in the $(R_0\neq 0)$-degenerate case, but rather that the only solution to equations \eqref{eq:D EOM dS} and \eqref{eq:D constraints dS} describing a localised perturbation is the one having $\mathcal{A}=0$, i.e.~the identically-vanishing solution, thus mirroring the $(R_0=0)$-degenerate case with $f''(0)\neq 0$.\footnote{
    Therefore, in $(R_0\neq 0)$-degenerate $f(R)$ models with $\fppo\neq 0$, the scalaron is once again not strongly-coupled \emph{stricto sensu}, yet no degrees of freedom propagate atop a dS background in said models.
} We can therefore state the ensuing Result:

\begin{result} \label{result:R_0 neq 0 D stability}
    At linear level in perturbations, there are no propagating degrees of freedom on top of a dS background space-time with $R=R_0$ in $(R_0\neq 0)$-degenerate $f(R)$ models. This is because, in said models, (i) the graviton is always strongly-coupled, (ii) the scalaron is strongly coupled as well provided that $\fppo=0$, and (iii) the only admissible solution for the scalaron perturbation is the vanishing one if $\fppo\neq 0$.
\end{result}

\section{Conclusions}
\label{Sec6}

In this communication, we have set out to determine the number of independent, gauge-invariant degrees of freedom propagating on a MS background space-time in $f(R)$ metric gravity, as well as their stability. To that end, we have first performed a model-independent perturbative expansion of the $f(R)$ EOM \eqref{EOM}, which we subsequently particularised for both types of $f(R)$ gravities admitting MS backgrounds: non-degenerate and degenerate. After isolating the gauge-invariant linear degrees of freedom propagating on top of the MS background in each case, we have solved the corresponding linearised EOM so as to assess the stability of their solutions.

The main findings of our investigation, encapsulated on Results \ref{result:DOF number ND}, \ref{result:graviton stability ND}, \ref{result:scalaron stability ND}, \ref{result:R_0 = 0 D stability} and \ref{result:R_0 neq 0 D stability} above, can be summarised as follows:
\begin{itemize}
    \item MS background solutions with scalar curvature $R=R_0$ are strongly-coupled in all $f(R)$ models such that $\fppo=0$, as the scalaron kinetic term disappears from the corresponding linearised EOM.
    \item $R_0$-non-degenerate models---i.e.~those such that $\fpo\neq 0$---additionally fulfilling $\fppo\neq 0$ propagate three gauge-invariant degrees of freedom on top of MS background solutions with $R=R_0$, two of which correspond to the usual massless and traceless graviton already found in GR, and one corresponding to the scalaron. Even though the graviton is always stable on these backgrounds, the scalaron can develop a tachyonic instability if its effective mass squared $m_\eff^2$---as given by \eqref{m2}---is negative.
    \item $R_0$-degenerate models---i.e.~those which have $\fpo=0$---do not possess stable MS backgrounds with $R=R_0$ even in cases where $\fppo\neq 0$, the reason being that, on the one hand, the graviton is always strongly-coupled, whereas, on the other hand, the only solution of the linearised scalaron EOM describing a small, localised perturbation is the identically vanishing one. Owing to this, it is clear that degenerate $f(R)$ models are not only incompatible with gravitational-wave observations, but also inherently pathological, as first pointed out in \cite{Casado-Turrion:2023rni}.
\end{itemize}

The results presented herein are consistent with the various recent studies focusing on the number of linear degrees of freedom in the subtle, purely-quadratic model $f(R)=\alpha R^2$ \cite{Alvarez-Gaume:2015rwa,Hell:2023mph,Golovnev:2023zen}, which initially provided some motivation for the present work. In fact, these earlier findings can all be obtained as particular instances of our more general results, which are valid for \emph{every} $f(R)$ model. More precisely, even though the purely-quadratic $f(R)$ model admits MS solutions having any constant scalar curvature $R=R_0$ as a consequence of scale invariance,\footnote{
    In fact, the model is restricted-conformal invariant, meaning that it is unnaffected by transformations of the form $\tilde{g}_{\mu\nu}=\Omega^2 g_{\mu\nu}$ such that $\Box\Omega=0$.
} it can easily be checked that the model is $(R_0\neq 0)$-non-degenerate but $(R_0=0)$-degenerate, with $\fppo\neq 0$ in both cases. Therefore, as per our general Results \ref{result:DOF number ND}, \ref{result:graviton stability ND} and \ref{result:scalaron stability ND}, the behaviour on MS background space-times with $R_0\neq 0$ is completely regular, with a massless and traceless graviton plus a massless scalar mode, both of which propagate stably. However, in the case of a Minkowski background, since $f'(0)=0$, Result \ref{result:R_0 = 0 D stability} above entails that the linearised spectrum around flat space-time is empty, in agreement with the existing literature. Because all the aforementioned previous works on the issue made use of techniques differing from standard perturbation theory (as done in this investigation), we confirm that different ways of deriving the spectrum of $f(R)=\alpha R^2$ gravity lead to exactly the same result, as expected. 

\section{Acknowledgments}

The authors would like to thank Jose Beltr\'{a}n Jim\'{e}nez, Alejandro Jim\'{e}nez Cano and Francisco Jos\'{e} Maldonado Torralba for their insightful comments and discussions. Funded by research grant PID2022-137003NB-I00 from Spanish MCIN/AEI/10.13039/501100011033/ and EU FEDER. ACT acknowledges support from a Universidad Complutense de Madrid-Banco Santander early-career researcher contract CT63/19-CT64/19, as well as from postdoctoral fellowship S-PD-24-135 of the Research Council of Lithuania. AdlCD acknowledges support from BG20/00236 action (MCINU, Spain), NRF Grant CSUR23042798041 (South Africa), CSIC Grant COOPB23096 (Spain), Project SA097P24 funded by Junta de Castilla y Le\'{o}n (Spain) and Grant PID2021-122938NB-I00 funded by MCIN/AEI/10.13039/501100011033 and by \textit{ERDF A way of making Europe}.

\section*{Appendices}
\appendix

\section{Gauge-fixing}
\label{Appendix: gauge-fixing}

In order to simplify equations \eqref{eq:ND linearised EOM} above even further, as well as to acquire a more precise idea of their physical significance, it is convenient to choose appropriate gauge-fixing conditions on $h_{\mu\nu}$. As it is well known, by performing a coordinate transformation
\begin{equation}
    x^\mu \rightarrow x_{(\xi)}^\mu= x^\mu + \xi^\mu,
\end{equation}
the metric perturbation $h_{\mu\nu}$ changes as
\begin{equation}
    h^{(\xi)}_{\mu\nu}=h_{\mu\nu}+\delta h_{\mu\nu},
\end{equation}
with
\begin{equation} \label{delta_hmunu}
    \delta h_{\mu\nu}=2\Do_{(\mu} \xi_{\nu)}.
\end{equation}
Therefore, its trace changes by
\begin{equation}
    \delta h= 2 \Do^\mu \xi_\mu.
\end{equation}
After some algebra, it is also possible to show that the first-order perturbation of the Ricci scalar is gauge-invariant, i.e.
\begin{equation} \label{delta_Rh}
    \delta R^\Rh = 0.
\end{equation}
Thus, combining \eqref{eq:barh} and \eqref{delta_hmunu}--\eqref{delta_Rh}, one finds that $\barh_{\mu\nu}$ changes by
\begin{equation}
    \delta \barh_{\mu\nu}=2\Do_{(\mu} \xi_{\nu)}
    -g^\szero_{\mu\nu}\Do^\rho \xi_\rho,
\end{equation}
from where one can obtain
\begin{equation}
    \Do^\mu\barh^{(\xi)}_{\mu\nu}=\Do^\mu\barh_{\mu\nu}+( \Box + \Lambda) \xi_\nu.
\end{equation}
Therefore, by choosing $\xi_\nu$ such that it satisfies
\begin{equation}
\label{BoxLambda_xi}
    (\Box + \Lambda) \xi_\nu = -\Do^\mu\barh_{\mu\nu},
\end{equation}
the new tensor $\barh^{(\xi)}_{\mu\nu}$ will fulfill
\begin{equation}
\Do^\mu\barh^{(\xi)}_{\mu\nu}=0.
\end{equation}
In the following, we will assume this generalization of the transverse (or de Donder) gauge condition, and therefore we will suppose that $\Do^\mu\barh_{\mu\nu}=0$. In such scenario, EOM \eqref{eq:ND linearised EOM} become
\begin{equation} \label{eq:ND linearised EOM T}
    \left(\Box-\dfrac{R_0}{6}\right)\barh_{\mu\nu}+\dfrac{R_0}{6}\barh g ^\szero_{\mu\nu} + \bigo(h^2) = 0.
\end{equation}
It is not difficult to show that the previous equation can also be written in the more compact form
\begin{equation}
 \Box \barh_{\mu\nu} + 2 R^\szero_{\mu\rho\nu\sigma}\barh^{\rho\sigma} + \bigo(h^2) = 0,
\end{equation}
whose trace constitutes an EOM for $\barh$:
\begin{equation} \label{eq:barh trace EOM}
    (\Box + 2\Lambda)\barh + \bigo(h^2) = 0. 
\end{equation}
The transverse condition $\Do^\mu\barh_{\mu\nu}=0$ does not fix the gauge completely and allows one to make further gauge transformations. In fact, the de Donder gauge condition is preserved whenever the new gauge parameters fulfil
\begin{equation} \label{eqn:GC}
    (\Box + \Lambda) \xi_\nu = 0. 
\end{equation}
Now, taking into account that
\begin{equation}
\delta \barh = - 2 \Do^\mu \xi_\mu,
\end{equation}
it is not difficult to see that it is possible to find a gauge transformation obeying both \eqref{eqn:GC} and
\begin{equation} \label{eqn:TGC}
    2 \Do^\mu \xi_\mu = \barh
\end{equation}
simultaneously. More precisely, the condition for this to happen is that the EOM \eqref{eq:barh trace EOM} for the trace $\barh$ in the transverse gauge shown above holds. Therefore, we can always use the gauge freedom to have the transverse $\Do^\mu\barh_{\mu\nu}=0$ and traceless $\barh=0$ conditions fulfilled at the same time, in what is known as the TT gauge. A straightforward consequence of the tracelessness of $\barh_{\mu\nu}$ is that, in TT gauge,
\begin{equation} \label{eq:TT gauge h}
    h= - \dfrac{4 \fppo}{\fpo} R^\Rh.
\end{equation}
Moreover, the fact that $\barh=0$ in TT gauge leads to the simplified EOM \eqref{EOM_h_tilde} upon substitution in \eqref{eq:ND linearised EOM T}.

Equation \eqref{EOM_h_tilde} cannot be simplified any further via gauge transformations. Nonetheless, as we shall show now, choosing the TT gauge does not exhaust the potential of gauge transformations to simplify the form of the tensor $\barh_{\mu\nu}$. In fact, parameters $\xi_\mu$ can always be split as
\begin{equation}
    \xi_\mu = \xi_\mu^\T + \Do_\mu \xi,
\end{equation}
whose transverse $\xi_\mu^\T$ part fulfils
\begin{equation}
    \Do^\mu\xi_\mu^\T=0,
\end{equation}
whereas its longitudinal part $\xi$ satisfies
\begin{equation}
    \Box \xi=\Do^\mu\xi_\mu.
\end{equation}
From equation \eqref{eqn:TGC}, we see that the longitudinal degree of freedom $\xi$ has already been employed to set the traceless condition on $\barh_{\mu\nu}$. Thus, if we wish to remain in the TT gauge, we must only consider purely transverse gauge transformations $\xi_\mu= \xi_\mu^\T$ respecting the following condition:
\begin{equation} \label{eqn:GC3}
    (\Box +\Lambda)\xi_\mu^\T=0.
\end{equation}
However, these restricted gauge transformations are enough to allow for the additional gauge choice
\begin{equation}
    \barh^{(\xi)}_{\mu0}=0.
\end{equation}
This entails that
\begin{equation}
\label{D_xi_0mu}
    \Do_\mu \xi_0^\T+\Do_0\xi_\mu^\T= -\barh_{\mu0},
\end{equation}
which is compatible with \eqref{eqn:GC3}, as it is easy to check. Now, by taking $\mu=0$ in \eqref{D_xi_0mu}, we have
\begin{equation}
    2\Do_0 \xi_0^\T= -\barh_{00},
\end{equation}
while, by taking $\mu=i=1,2,3$ in \eqref{D_xi_0mu}, one gets
\begin{equation}
\Do_i \xi_0^\T+\Do_0\xi_i^\T= -\barh_{i0}.
\end{equation}
As such, by solving these equations, it is possible in principle to find a gauge transformation so that the conditions in \eqref{eq:TT + h_mu0} are all fulfilled simultaneously.

At first sight, one could presume that there is a total of $4+1+4=9$ equations in \eqref{eq:TT + h_mu0}. However, only 8 of them are actually independent. This is because the transverse condition \smash{$\Do^\mu\xi_\mu^\T=0$} allows one to express \smash{$\Do_0\xi_0^\T$} in terms of \smash{$\Do_i\xi_i^\T$}. In addition, \smash{$\barh_{\mu\nu}$} is symmetric, and hence may in principle have up to 10 independent components, which are nonetheless related by the 8 conditions in \eqref{eq:TT + h_mu0}. Therefore, once gauge symmetry has been exhausted, one concludes that only two out of the ten components in $\barh_{\mu\nu}$ are truly gauge-independent. These two components correspond to the two degrees of freedom of the standard massless and traceless graviton found in GR.

\section{Mode solutions to the Klein-Gordon equation on dS space-time in planar coordinates}
\label{Appendix: mode solutions}

Let $\Phi$ be any field satisfying equation
\begin{equation} \label{eq:KG in dS}
    \left(-\partial_t^2 + \dfrac{\nabla^{2}}{a^2(t)}- 3 H_0 \partial_t-m^2\right) \Phi=0
\end{equation}
on dS space-time \eqref{dS}--\eqref{a_dS} in planar coordinates. For instance, $\Phi$ could be either the scalaron $R^\Rh$ of non-degenerate $f(R)$ models or the rescaled graviton $H_{ij}$ as defined in \eqref{eq: H_ij}, since their respective linearised field equations \eqref{dS scalar eqn ND} and \eqref{tensor_dalembert} are both of this form. If we seek separable solutions of the form
\begin{equation}
    \Phi_{\vec{p}}(t,\vec{x})=\phi_{\vec{p}}(t)\,\e^{\pm\iu\vec{p}\cdot\vec{x}},
\end{equation}
equation \eqref{eq:KG in dS} becomes
\begin{equation} \label{eq:KG in dS phi_p}
    \phi''_{\vec{p}}(t)+3H_0\phi'_{\vec{p}}(t)+\left(\dfrac{\vec{p}^{2}}{a^2(t)}+m^2\right)\phi_{\vec{p}}(t)=0,
\end{equation}
and there are two distinct possible scenarios that must be considered separately: \smash{$\vec{p}=\vec{0}$} and \smash{$\vec{p}\neq\vec{0}$}. This is because the limit of modes $\phi_{\vec{p}\neq\vec{0}}$ as \smash{$\vec{p}\rightarrow\vec{0}$} is not always well-defined, as we shall see in due course, cf.~footnote \ref{footnote:z}.

On the one hand, if \smash{$\vec{p}=\vec{0}$}, it is straightforward to check that \eqref{eq:KG in dS phi_p} turns into a damped harmonic oscillator equation for $\phi_{\vec{0}}$. As a result, the two independent solutions with vanishing wave vector are
\begin{equation}
    \phi^\sboth_{\vec{0}}(t,\vec{x})=\e^{-3 H_0 t/2}\,\e^{\pm\iu\omega_0 t},
\end{equation}
representing ingoing and outgoing damped plane waves of infinite wave-length and frequency
\begin{equation} \label{eq:omega_0}
    \omega_0=\sqrt{m^2-\Omega_0^2},
\end{equation}
where we have introduced
\begin{equation} \label{eq:Omega_0}
    \Omega_0\equiv\dfrac{3H_0}{2}.
\end{equation}
The complete zero-mode solution is thus given by
\begin{eqnarray} \label{eq:zero-mode}
    \Phi_{\vec{0}}(t,\vec{x}) & = & A^\sone(\vec{0})\,\e^{-3 H_0 t/2}\,\e^{+\iu\omega_0 t} \nonumber \\
    & + & A^\stwo(\vec{0})\,\e^{-3 H_0 t/2}\,\e^{-\iu\omega_0 t},
\end{eqnarray}
where amplitudes \smash{$A^\sboth(\vec{0})$} must be such that $\Phi_{\vec{0}}$ remains real.\footnote{
    We also remind the reader that the tensorial character of amplitudes \smash{$A^\sboth(\vec{0})$} depends on whether one is considering solutions to the scalaron equation \eqref{dS scalar eqn ND} or to the graviton equation \eqref{tensor_dalembert}. In the former case, the amplitudes will be scalars, whereas in the latter they will be purely-spatial, symmetric, transverse and traceless tensors; c.f.~\eqref{eq:ND graviton A_ij constraints 1} and \eqref{eq:ND graviton A_ij constraints 2}.
} If $\omega_0^2\leq 0$, this is always the case as long as both \smash{$A^\sboth(\vec{0})$} are real. However, if $\omega_0^2>0$, it is necessary to have \smash{$[A^\sone(\vec{0})]^*=A^\stwo(\vec{0})$}.

On the other hand, for \smash{$\vec{p}\neq\vec{0}$}, and in light of the results found above for the vanishing-wave-vector case, one can factorise $\phi_{\vec{p}}$ as
\begin{equation}
    \phi_{\vec{p}}(t)=\e^{-3 H_0 t/2}\, \psi_{\vec{p}}(t).
\end{equation}
Substituting this on \eqref{eq:KG in dS}, and introducing a new time coordinate $z\in(\infty,0)$,\footnote{
    By this we mean that $z\rightarrow\infty$ when $t\rightarrow-\infty$ and $z\rightarrow0$ when $t\rightarrow+\infty$, i.e.~time flows backwards when using $z$. The reason for this apparently unwieldy choice for the overall sign of $z$ (instead of the forward-flowing choice $z\rightarrow-z$) is that the mode solutions turn out to be multivalued, featuring a branch point at the origin and a branch cut along the real negative semi-axis. \label{footnote:z}
} defined by
\begin{equation} \label{eq:coord z}
    z\equiv\dfrac{|\vec{p}|}{H_0}\,\e^{-H_0 t},
\end{equation}
one finds that $\psi_{\vec{p}}$ satisfies Bessel's equation,
\begin{equation} \label{eq:Bessel}
    z^2\,\psi_{\vec{p}}''(z) + z\,\psi_{\vec{p}}'(z) + (z^2 - \nu^2)\, \psi_{\vec{p}}(z) = 0,
\end{equation}
where
\begin{equation} \label{eq:nu2}
    \nu^2\equiv-\dfrac{\omega_0^2}{H_0^2},
\end{equation}
where $\omega_0$ is given again by \eqref{eq:omega_0}--\eqref{eq:Omega_0}. Notice that index $\nu$ is real and positive for $\omega_0^2\leq 0$, but pure imaginary for $\omega_0^2>0$. Out of the various solutions to Bessel's equation \eqref{eq:Bessel}, the most suitable for the problem at hand are Hankel functions of the first and second kind \cite{Birrell:1982ix,Cotaescu:2008aq},
\begin{equation}
    \psi_{\vec{p}}(z)=\Hboth_\nu(z),
\end{equation}
which are analogous to the positive- and negative-frequency exponentials appearing on flat-space plane waves. Moreover, in the limit of large $z$, both Hankel functions reduce to damped plane waves:
\begin{equation}
    \Hboth_\nu(z)\underset{z\gg 1}{\sim}\sqrt{\dfrac{2}{\pi z}}\,\e^{\pm\iu(z-\pi\nu/2-\pi/4)}.
\end{equation}
As such, out of the four possible independent solutions to the mode equation with \smash{$\vec{p}\neq\vec{0}$}, namely \smash{$\Phi_{\vec{p}}\propto \Hboth_\nu \e^{\pm\iu\vec{p}\cdot\vec{x}}$}, there are only two leading to the correct plane-wave limit as $z\gg 1$. The complete mode solution in `planar-time' $t$ is then
\begin{eqnarray} \label{eq:k-mode}
    \Phi_{\vec{p}}(t,\vec{x}) & = & A^\sone(\vec{p})\,\e^{-3 H_0 t/2}\,\Hone_\nu\Left(\dfrac{|\vec{p}|}{H_0}\,\e^{-H_0 t}\Right)\,\e^{-\iu\vec{p}\cdot\vec{x}} \nonumber \\
    & + & A^\stwo(\vec{p})\,\e^{-3 H_0 t/2}\,\Htwo_\nu\Left(\dfrac{|\vec{p}|}{H_0}\,\e^{-H_0 t}\Right)\,\e^{+\iu\vec{p}\cdot\vec{x}}.\mybigskip\myskip
\end{eqnarray}
Given that
\begin{equation}
    [\Hboth_\nu(z)]^*=\Hrev_{\nu^*}(z^*),
\end{equation}
for $\nu>0$ (i.e.~if $\omega_0^2\leq 0$) it is necessary to require the amplitudes in \eqref{eq:k-mode} to satisfy \smash{$[A^\sone(\vec{p})]^*=A^\stwo(\vec{p})$} in order to have a real \smash{$\vec{p}$}-mode. For $\nu=\iu|\nu|$ (i.e.~if $\omega_0^2>0$), the condition to have a real mode $\Phi_{\vec{p}}$ is instead \smash{$[A^\sone(\vec{p})]^*=\e^{-\pi|\nu|}\,A^\stwo(\vec{p})$}.

\section{Considerations on models with $m_\eff^2<0$}
\label{Appendix: m2 < 0}

Concerning the physical interpretation of $m_\eff^2$---as given by \eqref{m2}---as the physical mass squared of the scalaron perturbation $R^{(h)}$, the following consideration is in order. As it is well known, whenever $f'(R)>0$, $f(R)$ theories can be formulated in the so-called Einstein frame. In order to do so, one performs the conformal transformation $\tilde g_{\mu\nu}= \Omega^2 g_{\mu\nu}$ given by
\begin{equation} \label{eq:f(R) conformal factor}
    \Omega^2 = f'(R) = \e^{\beta \phi}
\end{equation}
where $\beta^2\equiv 2 \kappa/3$ and $\phi$ is the Einstein-frame scalar field. As a result of the aforementioned transformation, the original $f(R)$ action becomes the pure EH action for the new metric $\tilde g_{\mu\nu}$ coupled to the canonically-normalised scalar field $\phi$, whose self-interactions are described by the potential
\begin{equation}
    V(\phi)=\dfrac{1}{2 \kappa}\dfrac{f'(R)R-f(R)}{f^{\prime 2}(R)}.
\end{equation}
Here, $R$ is understood to be a function of the scalar field $\phi$ through the conformal factor definition \eqref{eq:f(R) conformal factor} above. In this Einstein-frame description of the $f(R)$ theory, looking for constant-curvature solutions with $R=R_0$ is equivalent to seeking constant scalar-field configurations $\phi= \phi_0$ such that $\beta \phi_0 = \log f'(R_0) $. It is then clear that the stability of the solution requires $\phi_0$ to be a (local) minimum of the potential. In other words, two conditions must be met. First,
\begin{equation}
    \dfrac{\dif V(\phi)}{\dif\phi}=\dfrac{\beta}{2 \kappa}\dfrac{2f(R)-R f'(R)}{f^{\prime 2}(R)}
\end{equation}
must vanish at $\phi=\phi_0$ (i.e.~$R=R_0$), which implies that $2f(R_0)=R_0 f'(R_0)$. This is precisely the $f(R)$ trace EOM \eqref{eq:EOM trace MS} for constant-curvature solutions with $R=R_0$. Second,
\begin{equation}
    \dfrac{\dif^2 V(\phi)}{\dif\phi^2}=\dfrac{1}{3 f^{\prime 2}(R)}\left[\dfrac{f^{\prime 2}(R)}{f''(R)}+ R f'(R) -4f(R)\right]
\end{equation}
has to be positive at $\phi = \phi_0$. Using the trace EOM \eqref{eq:EOM trace MS} and the definition of the scalaron mass \eqref{m2}, it is trivial to check the following holds:
\begin{equation}
    \Left.\dfrac{\dif^2 V(\phi)}{\dif\phi^2}\Right|_{\phi=\phi_0}=\e^{-\beta \phi_0} m_\eff^2,
\end{equation}
where we remind the reader that, by construction, $f'(R)>0$ in order to have a well-defined Einstein frame representation. Therefore, we conclude that the stability condition for a MS solution with $R=R_0$ in $R_0$-non-degenerate $f(R)$ models amounts to condition $m_\eff^2(R_0)>0$. According to Result \ref{result:scalaron stability ND} in Section \ref{sec:scalaron ND}, this is equivalent to having a non-tachyonic scalaron, at least for those $f(R)$ models which can be consistently formulated in the Einstein frame.

\section{Discussion on the relationship between the Einstein-frame scalaron and the scalar perturbation $R^\Rh$}
\label{appendix:scalaron}

In non-degenerate $f(R)$ models such that $\fpo>0$ (so that the conformal transformation \eqref{eq:f(R) conformal factor} to the Einstein frame is well-defined), one can expand $f'(R)$ around $R_0$ as
\begin{equation}
    f'(R)=\fpo+\fppo\,R^{(h)}+\mathcal{O}(h^2),
\end{equation}
and it is also possible to expand the Einstein-frame scalaron---as defined through \eqref{eq:f(R) conformal factor}---to order $\mathcal{O}(h)$,
\begin{equation}
    \phi=\phi_0+\phi^{(h)}+\mathcal{O}(h^2).
\end{equation}
The linearised version of conformal transformation \eqref{eq:f(R) conformal factor} then entails
\begin{equation}
    f''(R_0)R^{(h)}=\beta f'(R_0)\phi^{(h)}.
\end{equation}
Thus, provided that $\fppo\neq 0$, one has that
\begin{equation}
    \phi^{(h)}=\dfrac{\fppo}{\beta\fpo}R^\Rh,
\end{equation}
i.e.~at linear order in perturbations, the Einstein frame scalaron is proportional to the Ricci scalar perturbation $R^\Rh$. Now, fixing the TT gauge, and taking into account equation \eqref{eq:TT gauge h} in Appendix \ref{Appendix: gauge-fixing}, we find
\begin{equation}
    h= - \dfrac{4 \fppo}{\fpo} R^\Rh=-4 \beta \phi^{(h)}.
\end{equation}
Therefore, at linear order, the scalars $h$, $R^\Rh$ and $\phi^{(h)}$ represent essentially the same quantity, the scalaron.

\bibliography{bibliography}

\end{document}